\documentclass[10pt,aps,prd,twocolumn,preprintnumbers,showpacs,nofootinbib,notitlepage,longbibliography,superscriptaddress]{revtex4-1}

\usepackage{amsmath, amssymb,braket}
\usepackage{slashed}
\usepackage{color}
\usepackage{float}
\usepackage{graphicx}
\usepackage{subfigure}
\usepackage{natbib}
\usepackage{tikz}
\usepackage[colorlinks=true,citecolor=blue,urlcolor=blue]{hyperref}
\usepackage{xspace}

\newcommand{\dbar}{d\hspace*{-0.08em}\bar{}\hspace*{0.1em}}

\begin{document}
\title{Inelastic Freeze-in}

\author{Saniya Heeba}
\email{saniya.heeba@mcgill.ca}
\affiliation{Department of Physics \& Trottier Space Institute, McGill University, Montr\'{e}al, QC H3A 2T8, Canada}

\author{Tongyan Lin}
\email{tongyan@physics.ucsd.edu}
\affiliation{Department of Physics, University of California, San Diego, CA 92093, USA}

\author{Katelin Schutz}
\email{katelin.schutz@mcgill.ca}
\affiliation{Department of Physics \& Trottier Space Institute, McGill University, Montr\'{e}al, QC H3A 2T8, Canada}

\begin{abstract}
	\noindent Dark matter (DM) could be a nonthermal relic that freezes in from extremely weak, sub-Hubble annihilation and decay of Standard Model (SM) particles. The case of Dirac DM freezing in via a dark photon mediator is a well-studied benchmark for DM direct detection experiments. Here, we extend prior work to take into account the possibility that DM is pseudo-Dirac with a small mass splitting. If the mass splitting is greater than twice the electron mass but less than the dark photon mass, there will be distinct cosmological signatures. The excited state $\chi_2$ is initially produced in equal abundance to the ground state $\chi_1$. Subsequently, the excited state population decays at relatively late cosmological times, primarily via the three-body process $\chi_2 \rightarrow \chi_1 e^+ e^-$. This process injects energetic electrons into the ambient environment, providing observable signatures involving Big Bang nucleosynthesis, cosmic microwave background spectral distortions and anisotropies, and the Lyman-$\alpha$ forest. Furthermore, the ground state particles that are populated from the three-body decay receive a velocity kick, with implications for DM clustering on small scales. We find that cosmological probes and accelerator experiments are highly complementary, with future coverage of much of the parameter space of the model. 
\end{abstract}

\maketitle

\section{Introduction}
\label{sec:intro}

The origins of dark matter (DM) in the early Universe remain an open question, with many possible thermal and nonthermal channels leaving a relic DM abundance that matches the observed present-day quantity $\Omega_c h^2 = 0.1200 \pm 0.0012 $~\cite{Planck:2018vyg}. If DM has a very small coupling to the Standard Model (SM) plasma such that interactions always occur at a rate below the Hubble expansion rate, then DM could be produced by the freeze-in mechanism. In this case, the DM is not in thermal equilibrium with the SM, and its density accumulates via rare annihilations and decays of particles in the SM thermal bath~\cite{Hall:2009bx,Chu:2011be,Bernal:2017kxu}. Many DM candidates, including sterile neutrinos and certain supersymmetric DM candidates, are naturally produced by the freeze-in mechanism~\cite{Dodelson:1993je,Moroi:1993mb,McDonald:2001vt,Covi:2002vw,Asaka:2005cn,Asaka:2006fs, Gopalakrishna:2006kr, Kusenko:2006rh,Page:2007sh,Petraki:2007gq,Cheung:2011mg,Cheung:2011nn,Chu:2013jja,Bae:2014rfa, Kolda:2014ppa,Shakya:2015xnx,Monteux:2015qqa,Co:2015pka,Co:2016fln,Benakli:2017whb}. Freeze-in is a rather general mechanism that can be realized in many models and produce the observed abundance of DM, provided the appropriate range of couplings~\cite{Asadi:2022njl}. 

The freeze-in mechanism naturally complements the possibility of a dark sector, where DM exists alongside other auxiliary particle content beyond the SM. For example, if there is a mediator that has a small coupling both to the SM and to the DM, then the small effective freeze-in couplings can be readily explained as originating from the product of two small numbers. A simple, technically natural example of such a mediator is a kinetically mixed dark photon~\cite{Holdom:1985ag}. Freeze-in of Dirac fermions via dark photon interactions has emerged as a key benchmark model for direct detection~\cite{Battaglieri:2017aum,Knapen:2017xzo,Dvorkin:2019zdi} as well as cosmological searches for DM~\cite{Dvorkin:2020xga}. Moving beyond this simple scenario, we note that when the dark gauge symmetry is broken, the Dirac fermions can split into two Majorana mass eigenstates and the couplings in this DM model are purely off-diagonal. This allows for inelastic processes that can be exothermic or endothermic, leading to markedly different behavior from the pure Dirac counterpart. 

The phenomenology of DM with inelastic interactions has been studied extensively, originally being proposed for its novel direct detection and astrophysical signatures~\cite{Tucker-Smith:2001myb,Finkbeiner:2007kk,Finkbeiner:2008gw, Arkani-Hamed:2008hhe, Cheung:2009qd, Chen:2009ab, Batell:2009vb,Graham:2010ca}. For pseudo-Dirac DM, the mechanisms for producing the relic abundance include the standard freeze-out production mechanism~\cite{Baryakhtar:2020rwy,CarrilloGonzalez:2021lxm} as well as the resonant~\cite{Feng:2017drg} and forbidden~\cite{Fitzpatrick:2021cij} regimes. This candidate has also recently been studied in the freeze-in regime~\cite{An:2020tcg} for $\sim$MeV-scale mediators and $\sim$keV-scale mass splittings motivated by the XENON1T excess~\cite{XENON:2020rca}, though this was subsequently excluded by XENONnT~\cite{XENON:2022ltv}.

In this paper, we focus on freeze-in of pseudo-Dirac DM via dark photon mediators. We work in the limit where the mediator is lighter than the DM, $m_{A'} < m_\chi$, which renders $m_{A'}$ as an irrelevant scale during freeze-in, making the required couplings insensitive to the mediator mass. Additionally, we will consider the regime where the $A^\prime-$DM coupling is smaller than the $A^\prime-$SM coupling, $g_\chi \lesssim e \epsilon$. Then the DM is dominantly produced through annihilations and decays of SM particles rather than freezing-in from a thermally populated dark sector. Freeze-in from the SM poses the most predictive scenario, which we shall subsequently refer to as ``visible freeze-in" and which we will show can be targeted by collider searches for the mediator in the $m_{A'} \sim 10\ {\rm MeV} - 10\ {\rm GeV}$ mass window. 

Motivated by these considerations as well as by the possibility of complementary late-Universe phenomenology, we primarily consider DM in the mass range $m_\chi \sim 30 \,\mathrm{MeV}- 1\,\mathrm{ TeV}$ with mass-splittings $\delta > 2 m_e$. With $\delta> 2 m_e$, the standard lore (with some exceptions, e.g.~\cite{Baryakhtar:2020rwy,CarrilloGonzalez:2021lxm}) for pseudo-Dirac DM in the freeze-out regime is that the excited state is efficiently depleted thermally in the early Universe. However, due to the small couplings in the freeze-in regime, the excited state does not necessarily get thermally de-populated and therefore there can be a metastable relic abundance of the excited state. 
We focus on the regime $\delta<m_{A'}$, which prevents the $\chi_2$ from decaying rapidly via $\chi_2 \rightarrow \chi_1 A'$.
For $\delta > 2 m_e$, there will then be a late decay of the excited state to the ground state plus charged SM leptons, $\chi_2 \rightarrow \chi_1 \ell^+ \ell^-$. Meanwhile, for $\delta < 2 m_e$ the lifetime of $\chi_2$ would be many orders of magnitude longer than the age of the Universe since decays to neutrinos are suppressed by $G_F^2$ and phase space factors, and $\chi_2 \rightarrow \chi_1 + 3 \gamma$ is also suppressed by powers of $\alpha_{em}$ and phase space factors \cite{Fitzpatrick:2021cij}. For the couplings relevant to freeze-in, the decay $\chi_2 \rightarrow \chi_1 \ell^+ \ell^-$ can occur during cosmological epochs where the injection of energetic charged leptons can heat or ionize the baryons, causing an observable departure from the standard cosmology. Furthermore, the ground-state DM particle produced in the decay receives a velocity kick that would alter the formation of structure on small scales, which could be observable with probes of cosmological clustering.

The rest of this paper is organized as follows. In Section~\ref{sec:model}, we review the model of pseudo-Dirac DM coupled to a kinetically mixed dark photon. We establish self-consistent theoretical motivations for considering our chosen parameter space that are complementary to the more phenomenological motivations described above. 
In Section~\ref{sec:early}, we compute the freeze-in relic abundance, further establishing that the 50\% of DM produced initially in the excited state is not immediately depleted by scattering processes. Specifically, we confirm that both $\chi_2 f \to \chi_1 f$ and $\chi_2 \chi_2 \to \chi_1 \chi_1$ are inefficient in the early Universe, ensuring a rich cosmological scenario where the excited state is metastable and can decay at later times. In Section~\ref{sec:decay} we compute the differential decay width for the three-body decay, finding that most of the energy from the mass splitting ends up in the electrons and finding that the $\chi_1$ gets a velocity kick of order $v \sim \delta/m_\chi$, resulting in half of the DM being ``warmed up" after the excited state decays. In Section~\ref{sec:cosmo}, we explore existing cosmological constraints and future probes of the decay of metastable excited states from freeze-in (from both the injected electrons and the recoiling $\chi_1$). Discussion and concluding remarks follow in Section~\ref{sec:conclusions}.

\section{Model and Parameter Space}
\label{sec:model}
We consider a pseudo-Dirac DM model with two nearly mass-degenerate Majorana states $\chi_1$ and $\chi_2$, where the mass splitting is represented by $\delta = m_{\chi_2}- m_{\chi_1}$. This dark sector is coupled to the Standard Model through a massive dark photon, $A^\prime$, with dark gauge coupling $g_\chi$. The dark photon kinetically mixes with the SM hypercharge with mixing strength $\epsilon$, with the interaction Lagrangian, 
\begin{equation}
\label{eq:interactionL}
  \mathcal{L}\supset -\frac{\epsilon}{2}F^\prime_{\mu\nu}F_Y^{\mu\nu} - i g_\chi A^\prime_\mu \bar{\chi_2} \gamma^\mu \chi_1 .
 \end{equation}
Rotating the fields to a basis with canonical kinetic terms, below the scale of electroweak symmetry breaking we obtain \cite{Evans:2017kti,Hoenig:2014dsa} 
\begin{align}
  & \mathcal{L} \supset -i\left(g_\chi A^\prime_\mu + g_\chi \epsilon \sin \theta_W \frac{m_Z^2}{m_Z^2-m_{A^\prime}^2}Z_\mu\right) J^\mu_\mathrm{DM} \\
  & + \frac{e \epsilon}{\cos\theta_W}\frac{m_{A^\prime}^2}{m_Z^2 - m_{A'}^2}A_\mu^\prime J^\mu_{Y} -e\epsilon \cos\theta_W \frac{m_Z^2}{m_Z^2 - m_{A'}^2} A^\prime_\mu J^\mu_\mathrm{EM}\,, \nonumber
\end{align}
where $m_{A'}$ is the dark photon mass, $J^\mu_{\mathrm{EM}(Y)}$ is the electromagnetic (hypercharge) SM current and $J^\mu_\mathrm{DM} = \bar{\chi}_2\gamma^\mu\chi_1$ is the off-diagonal DM current. 
\subsection{Origin of dark sector masses and interactions}

As a minimal setup for generating the dark sector masses and interactions, we consider a complex dark Higgs field, $\Phi_D$, with a dark charge of two that couples to a standard Dirac fermion $\Psi$ with interactions \cite{Duerr:2020muu, Elor:2018xku}
\begin{align}
  \mathcal{L} = & \, i\bar{\Psi}\slashed{D}\Psi - m_D\bar{\Psi}\Psi + (D_\mu\Phi_D)^\dagger (D^\mu\Phi_D) \nonumber \\
  & - y_\chi(\bar{\Psi}^C\Psi \Phi_D^\ast + \mathrm{h.c}) + V(\Phi_D)\,,
\end{align}
where $D_\mu \equiv \partial_\mu - i q g_\chi A^\prime_\mu$ is the covariant derivative, $q = 1(2)$ is the charge for $\Psi(\Phi_D)$, $V(\Phi_D)$ is the dark Higgs potential and $C$ denotes charge conjugation. The vev of the dark Higgs, $\Phi_D = (v_D + h_D)/\sqrt{2}$ breaks the $U(1)_D$ symmetry, generating the dark photon mass. Additionally, the Yukawa interaction results in a Majorana mass term for the two components of $\Psi$. The corresponding fermion mass matrix is given by
\begin{align}
\frac{1}{2}(\eta \quad \xi)\left(\begin{matrix}
m_M & m_D \\
m_D & m_M
\end{matrix}\right)
\left(\begin{matrix}
\eta \\
\xi
\end{matrix}\right) + \mathrm{h.c.}\,,
\end{align}
with $m_M = \sqrt{2} y_\chi v_D$. This can be diagonalised to obtain two nearly mass-degenerate states, $\chi_1 = i(\eta -\xi)/\sqrt{2}$ and $\chi_2 = (\eta + \xi)/\sqrt{2}$. This change of basis gives the off-diagonal $A'$ interaction in Eq.~\ref{eq:interactionL}. In addition, there are interactions involving the dark Higgs, but we will be working in the limit of a heavy dark Higgs such that the interactions do not play an important role in the phenomenology.

In this scenario, we have the following relationships in the full theory: 
\begin{align}
\label{eq:UV_theory}
  m_{A^\prime} &= 2 g_\chi v_D, \quad 
  \delta = 2 \sqrt{2}y_\chi v_D \nonumber \\
  m_{\chi_{1,2}} &= m_D \mp \sqrt{2} y_\chi v_D \quad
  m_{h_D} = \sqrt{2 \lambda_D} v_D\,.
\end{align}
Here $\lambda_D$ is the dark Higgs quartic coupling. The splitting and overall mass scale of the DM depends on the relative size of the Majorana mass $\sqrt{2}y_\chi v_D$ to the Dirac mass $m_D$. Throughout the paper, we will generally work in the limit of $\delta \ll m_{\bar{\chi}},$ where $m_{\bar{\chi}} = (m_{\chi_1} + m_{\chi_2})/2$ can be considered as an average mass scale.

\subsection{Parameter Space}

In this work, we will be primarily interested in the dark photon mass range $m_{A^\prime} \in [10 \, \mathrm{MeV},\, 10\,\mathrm{GeV}]$, a benchmark for various ongoing and upcoming terrestrial searches (see for instance Fig. \ref{fig:acc_bounds}). Since we will work in the light mediator limit, $m_{A^\prime}< m_{\bar{\chi}}$, we will consider $m_{\bar{\chi}} \sim 30\, \mathrm{MeV}- 1\,\mathrm{TeV}$. For $m_{\bar{\chi}} \lesssim 30 \,\mathrm{MeV}$, strong bounds on $m_{A^\prime}$ coming from accelerator searches imply that the light mediator limit is difficult to satisfy, whereas for $m_{\bar{\chi}} \gtrsim 1\,\mathrm{TeV}$, the Electroweak Phase Transition may affect freeze-in production, which is beyond the scope of what we consider in this work. The parameter space we consider is also compatible with astrophysical constraints on dark photons from stellar energy loss~\cite{An:2013yfc}, SN1987A~\cite{Chang:2016ntp}, and cosmological constraints from Big Bang Nucleosynthesis (BBN)~\cite{Redondo:2008ec}.

Our focus is on visible freeze-in, where a single combination of couplings, $\epsilon g_\chi$, determines both the DM relic abundance and late-Universe phenomenology. As we will show in the next Section, in order to obtain the DM relic abundance in this scenario the typical couplings required are $\epsilon g_\chi \sim 10^{-11}$. For visible freeze-in, we also require $g_\chi \lesssim e\epsilon$, which implies small dark gauge couplings $g_\chi\lesssim 10^{-6}$. In the region of parameter space where $g_\chi$ is larger than $\epsilon e$, the production of DM would primarily be through dark photon fusion, which scales as $g_\chi^4$, and destroys this tight connection between early-universe production and late-universe observables. Larger values of $g_\chi$ could also overproduce DM through dark photon fusion, although this requires further study accounting for in-medium effects~\cite{An:2013yfc}. Furthermore, keeping $g_\chi$ small is more interesting cosmologically because this prevents the depopulation of the excited state through $\chi_2\chi_2\to\chi_1\chi_1$, as discussed in the next Section. Motivated by this potential for a richer cosmology, we also focus on the parameter space where $\delta < m_{A^\prime}$, which ensures that the excited state is cosmologically long-lived.

Finally, in order to isolate the effects of the pseudo-Dirac DM freeze-in, we require the dark Higgs to be sufficiently heavy to not meaningfully participate in any freeze-in dynamics or late-time interactions, with $m_{h_D} \gg m_{\bar{\chi}},\,m_{A^\prime}$. This requirement also has the effect of ensuring that in our parameter space, the dark Higgs is so heavy as to not be subject to strong stellar bounds~\cite{An:2013yua}. Since the heaviest DM we consider has a mass of 1 TeV, $m_{h_D} \gg 1\,\mathrm{TeV}$. Together with the relationships in Eq. \ref{eq:UV_theory}, $m_{h_D} \lesssim m_{A^\prime}/g_\chi$, this suggests that $m_{A^\prime},\,m_{\bar{\chi}} \ll m_{h_D}$ over our entire parameter space for $g_\chi\lesssim 10^{-6}$, consistent with our assumptions.

\section{Early Universe Behaviour}
\label{sec:early}
\subsection{Freeze-in production of DM}
For $g_\chi \lesssim e\epsilon$, the dominant channels contributing to freeze-in are the annihilation of SM fermions, $f\bar f\to\chi_1\chi_2$, or $Z$-decays, $Z\to\chi_1\chi_2$. The latter process is most important in the mass range GeV~$\lesssim 2m_{\bar{\chi}} < m_Z$. As noted in the previous Section, dark photon fusion is a negligible DM production channel for the range of couplings we consider. We additionally note that purely finite-temperature production channels, such as plasmon decay to $\chi_1 \chi_2$, are negligible for DM masses above $\sim 1$~MeV~\cite{Dvorkin:2019zdi}.We also ignore the RG running of the couplings and the temperature corrections to the processes considered here. For production via $Z$-decays that happens at $T\sim m_Z/3$, thermal corrections to the $Z$-mass are much smaller than the vacuum value of the $Z$-mass~\cite{PhysRevD.55.6253}. For heavier DM, thermal corrections (e.g. thermal masses) may have an O($10\%$) effect on the DM abundance~\cite{Bringmann:2021sth} but will not change our results qualitatively.

\begin{figure*}
\includegraphics[width=0.49\textwidth]{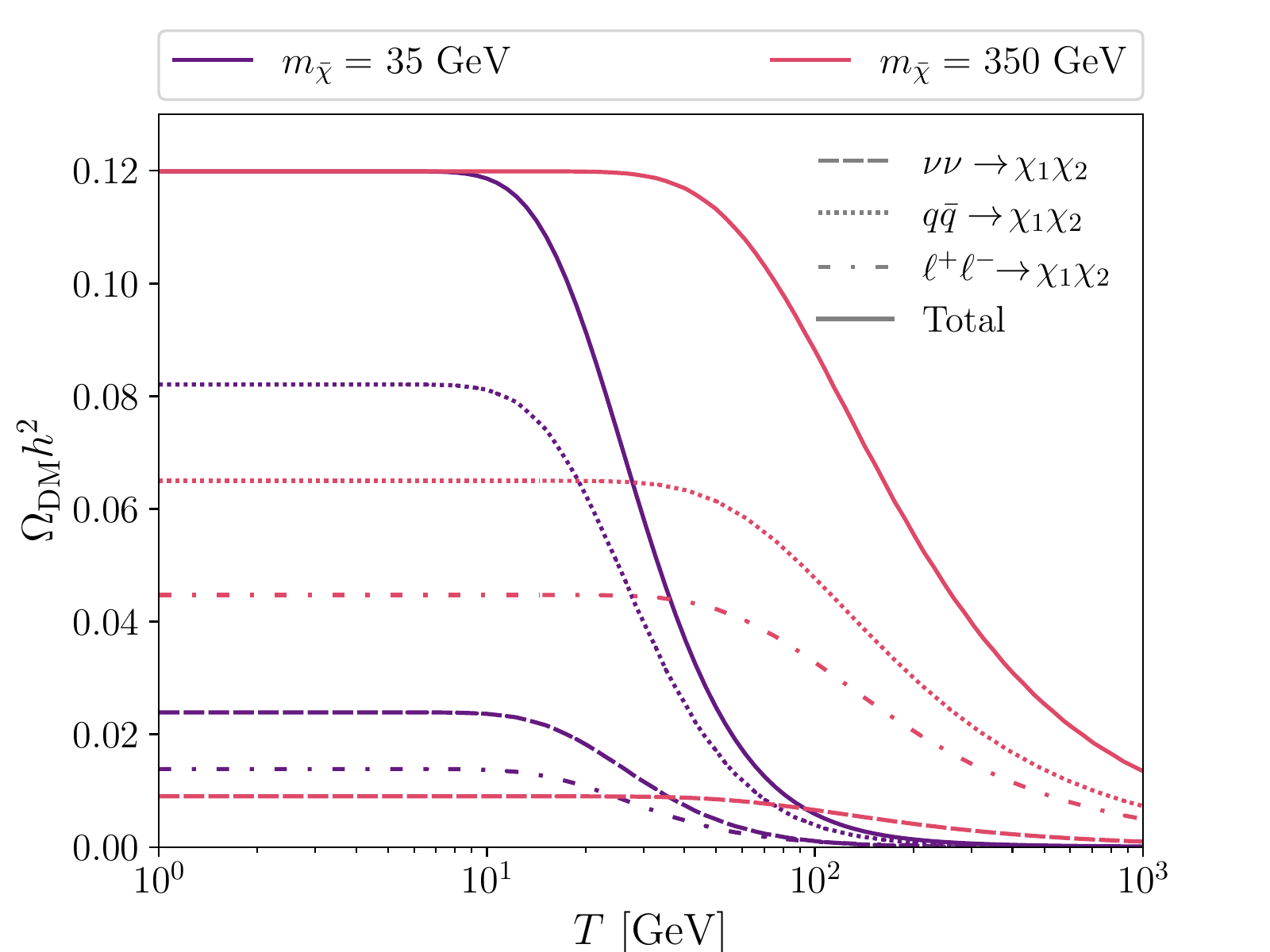}
  \includegraphics[width=0.49\textwidth, trim={0 15 0 14}, clip]{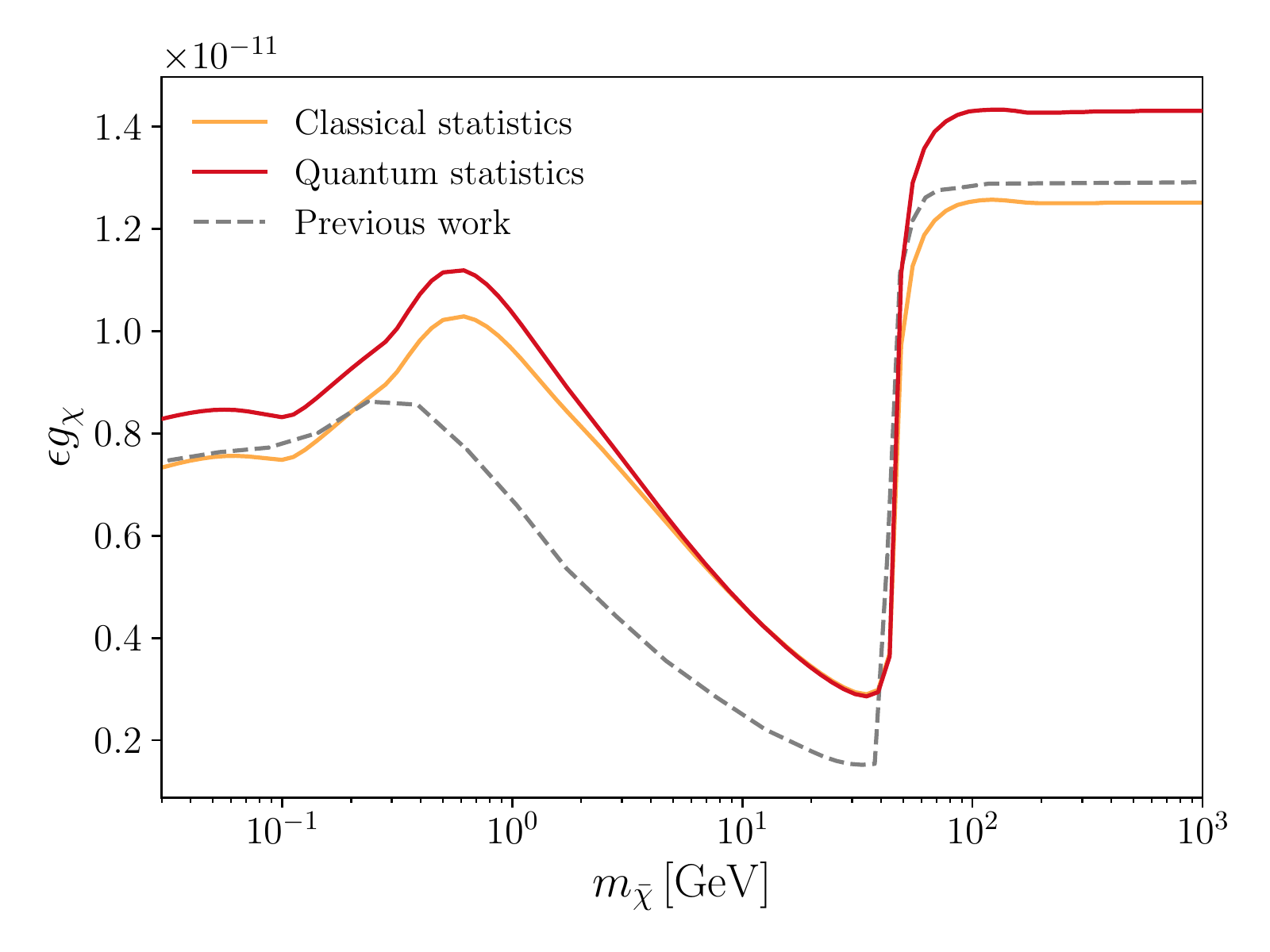}
  \vspace{-0.3cm}
  \caption{Left: Evolution of the various contributions to the total DM abundance today, including the neutrino (dashed), quark (dotted) and charged lepton (dot-dashed) channels for $m_{\bar{\chi}} = 35\,\mathrm{GeV}$ (purple) and $m_{\bar{\chi}} = 350\,\mathrm{GeV}$ (pink). The masses are chosen to yield representative values of the freeze-in couplings that are maximally or minimally affected by $Z$-decays. Right: The coupling combination that reproduces the DM relic density in the light mediator limit. The result is shown as a function of the DM mass assuming classical (yellow) and quantum (red) statistics. The dashed grey line corresponds to results from previous work \cite{Chu:2011be}. These results are independent of $m_{A^\prime}$ and $\delta$ assuming $m_{A^\prime} \ll m_{\bar \chi}$, $m_{A^\prime} \ll m_Z$, and $ \delta \ll m_{\bar \chi}$.
  }
  \vspace{-0.3cm}
  \label{fig:freezein_abundance2}
\end{figure*}

Given the assumptions above, the DM number density can be obtained by solving the Boltzmann equation,
\begin{equation}
\label{eq:boltzmann}
 sHx \frac{ \mathrm{d}Y_{\chi_1}}{\mathrm{d}x} =\langle\sigma v\rangle_{f \bar f\to\chi_1\chi_2} n_f^2 + \langle \Gamma\,\rangle_{Z\to\chi_1\chi_2}  n_Z\,,
\end{equation}
where $Y_{\chi_1} = n_{\chi_1}/s$ is the comoving number density of the ground or excited state, $H$ is the Hubble rate, $s$ is the entropy density, $x = m_{\bar{\chi}}/T$, and the two terms on the right correspond to the thermally-averaged annihilation and decay rates, respectively.

For $\delta \ll m_{A^\prime}$ and $T\gg m_{\bar{\chi}}$, the thermally averaged annihilation cross section above scales as $e^2\epsilon^2g_\chi^2\cos^2\theta_W/T^2$ in the light mediator limit, whereas the decay rate scales as $\epsilon^2g_\chi^2 \sin^2\theta_W m_Z^2/(2 T)$ for $T\gg m_Z$. The corresponding contributions to the total DM comoving number density can then be obtained by integrating Eq. \ref{eq:boltzmann} from $x\to0$ to $x\to\infty$, giving
\begin{align}
\label{eq:Ychi}
    Y_{\chi_1}^{ff} &\propto \frac{e^2\epsilon^2g_\chi^2 \cos^2\theta_W M_{\mathrm{Pl}}}{m_{\bar{\chi}}}
    \\
    Y_{\chi_1}^{Z} &\propto \frac{\epsilon^2g_\chi^2 \sin^2\theta_W M_{\mathrm{Pl}}}{m_Z}
\end{align}
where $M_{\mathrm{Pl}}$ is the Planck mass. 

For a more precise calculation of the DM production rate that includes the relevant relativistic spin statistics factors for particles in the initial state, we use the freeze-in routines of \texttt{micrOMEGAs}~\cite{Belanger:2018ccd} with a pseudo-Dirac DM model implemented in \texttt{CalcHEP}~\cite{Belyaev:2012qa}. The resulting DM production rate through various channels is plotted in the left panel of Fig. \ref{fig:freezein_abundance2} for $m_{\bar{\chi}} = 35\,\mathrm{GeV}$ and $m_{\bar{\chi}} = 350\,\mathrm{GeV}$. 
Note that \texttt{micrOMEGAs} includes the contribution from decays within the corresponding $2\to2$ processes in which the decaying particle can be produced on-shell (see Ref.~\cite{Belanger:2018ccd} for details). This can be seen in the left panel of Fig. \ref{fig:freezein_abundance2} as the difference in slope of the total production rate for $m_{\bar{\chi}} = 35\,\mathrm{GeV}$, which gets a significant contribution from $Z$-decays, versus 
$m_{\bar{\chi}} = 350\,\mathrm{GeV}$, which does not.

Summing up the various contributions gives us the total DM abundance,
\begin{equation}
\label{eq:omegah}
  \Omega_\mathrm{DM} = \frac{(Y_{\chi_1}^0 m_{\chi_1} +Y_{\chi_2}^0 m_{\chi_2} )s^0}{\rho_c^0}\, \approx \frac{2Y_{\chi_1}^0 m_{\bar{\chi}}s^0}{\rho_c^0}\,.
\end{equation}
Here, $\rho_c^0$ and $s^0$ are the critical energy density and entropy density today. The last approximation comes from the fact that the ground and excited states are symmetrically produced because for $\delta \ll m_{\bar \chi}$, the model effectively behaves as standard Dirac DM at the time of freeze-in when $T \gtrsim m_{\bar\chi}$. Requiring that the final abundances matches the observed relic density \cite{Planck:2018vyg}
gives us the product of the couplings $g_\chi \epsilon$ as a function of the DM mass. The final result, $g_\chi\epsilon \sim O(10^{-11})$ in the DM mass range of interest, is plotted in Fig. \ref{fig:freezein_abundance2} (right). For $2 m_{\bar \chi} > m_Z$, $g_\chi\epsilon$ is independent of the DM mass, which can also be inferred from Eqs.~\ref{eq:Ychi}-\ref{eq:omegah}.
We note that our results agree with those of \cite{Chu:2011be} above a few hundred GeV and below 100 MeV, but differ by an $\mathcal{O}(1)$ factor around the $Z$ resonance (dashed line in Fig. \ref{fig:freezein_abundance2}). This can be traced to a missing factor of $\tan\theta_W$ in Eq.~(73) of \cite{Chu:2011be}.

\begin{figure*}[t]
  \includegraphics[width=0.49\textwidth]{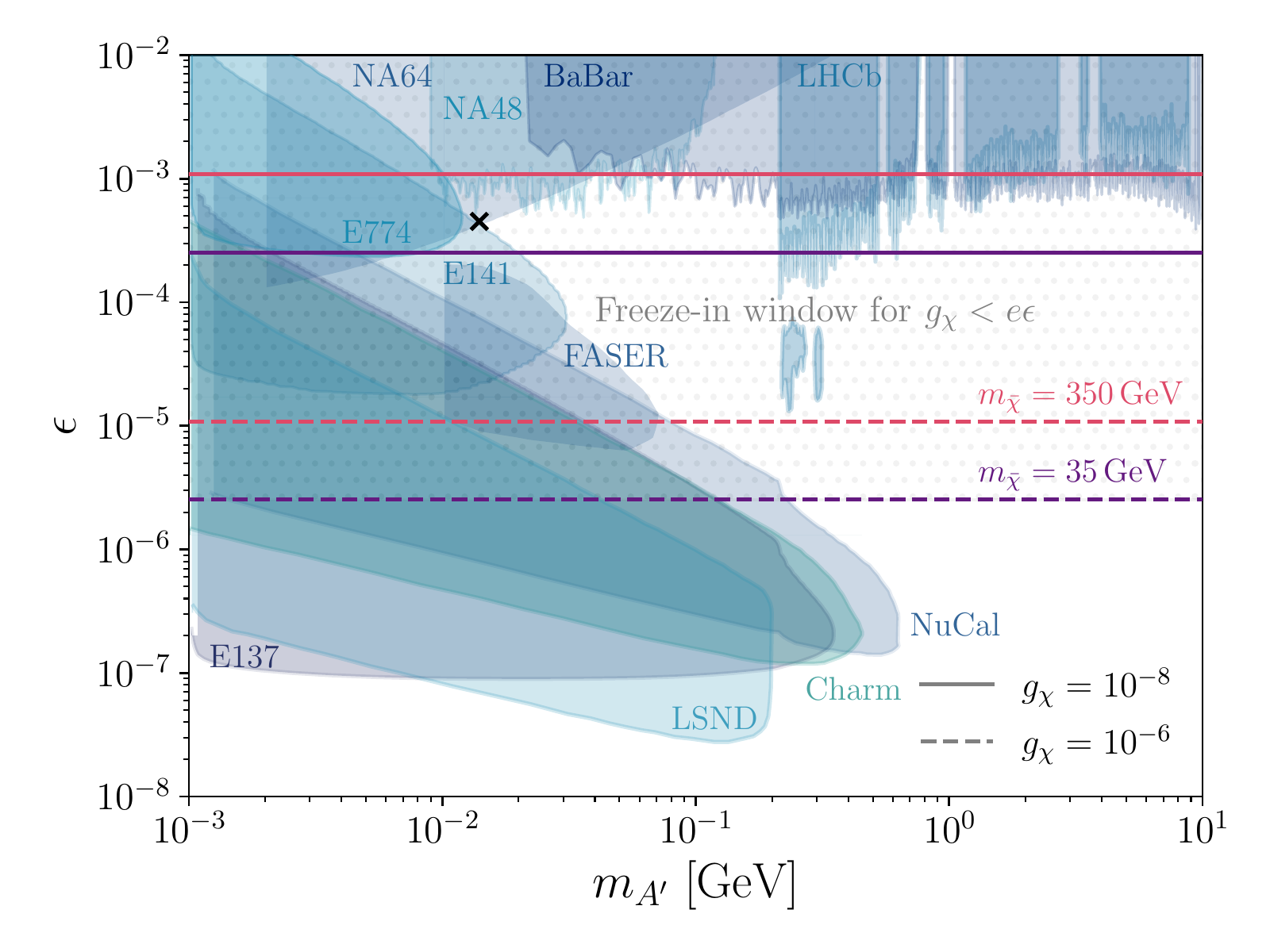}
  \includegraphics[width=0.49\textwidth]{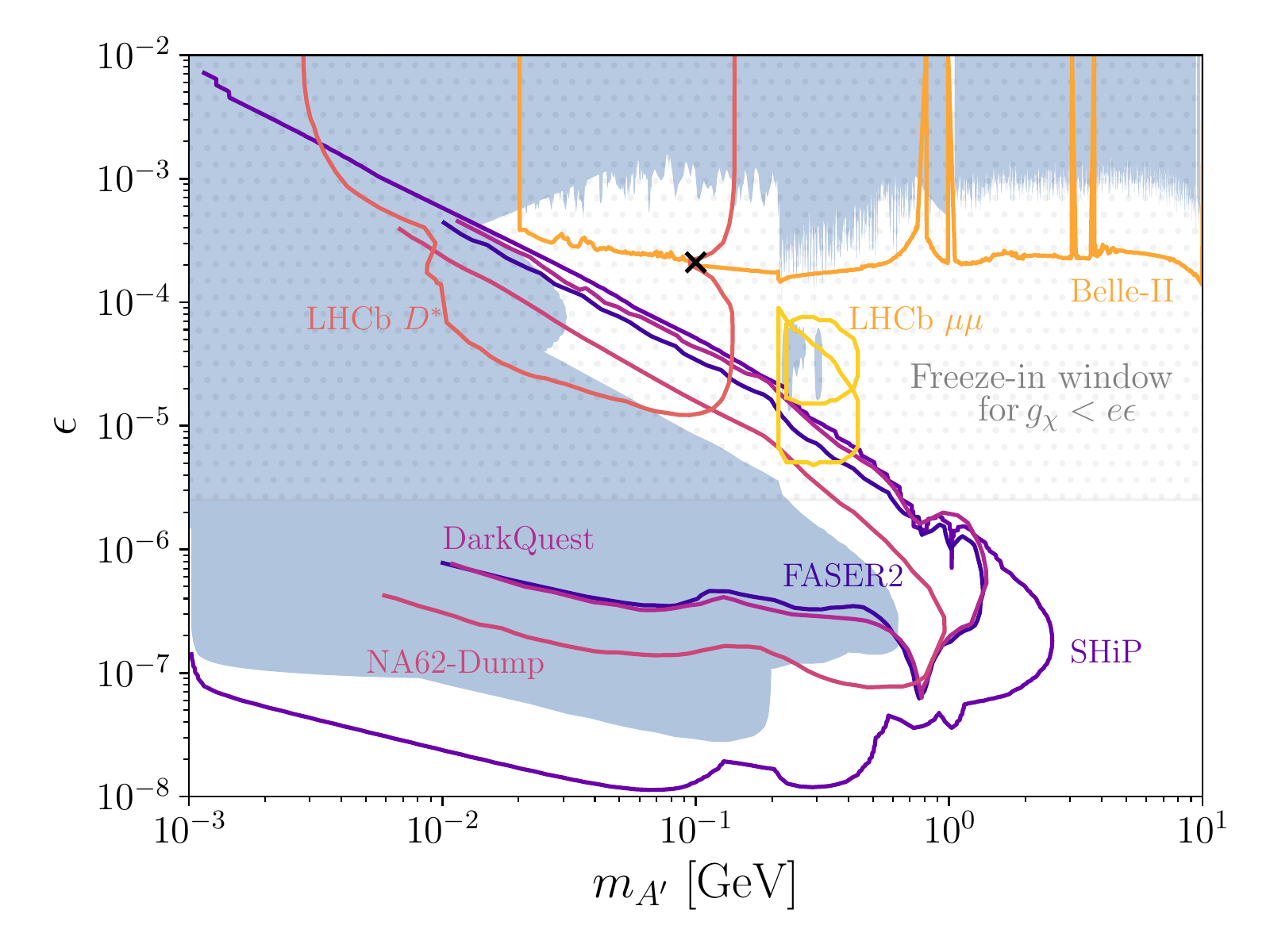}
  \caption{Left: Bounds on the freeze-in parameter space (dotted region) from beam dump and collider experiments. See \cite{Bauer:2018onh,Agrawal:2021dbo,CERN-FASER-CONF-2023-001} for a review. The horizontal lines show the values of $\epsilon$ for which we obtain the observed abundance of DM for $g_\chi = 10^{-8}$ (solid) and $g_\chi=10^{-6}$ (dashed) and $m_{\bar{\chi}} = 35 \,\mathrm{GeV}$ (purple) and $m_{\bar{\chi}} = 350 \,\mathrm{GeV}$ (pink). In this paper, we focus on the window where $g_\chi < e \epsilon$ which gives visible freeze-in from SM particles (as opposed to from the dark sector). Right: Projections from various experiments. 
  }
  \label{fig:acc_bounds}
\end{figure*}

In Fig.~\ref{fig:acc_bounds}, we compare our freeze-in parameter space with accelerator searches for dark photons in the $\epsilon-m_{A^\prime}$ plane. Assuming $g_\chi < 10^{-6}$, there is a minimum kinetic mixing for freeze-in to produce the observed relic abundance. This minimum on the freeze-in window is given by the dashed lines for $m_{\bar{\chi}} = 35\,\mathrm{GeV}$ and $m_{\bar{\chi}} = 350\,\mathrm{GeV}$. As noted above, for DM with masses around $\sim10$~GeV, the primary mode of production occurs via $Z$-decays, so smaller values of $\epsilon g_\chi$ produce the observed relic DM density compared to the $m_{\bar{\chi}} = 350\,\mathrm{GeV}$ benchmark. Additionally, for $m_{A^\prime}, \delta \ll m_{\bar{\chi}}$, the DM freeze-in abundance is independent of the dark photon mass and the mass-splitting. The shaded regions of Fig.~\ref{fig:acc_bounds} show bounds (left) and projections (right) from various fixed-target, beam dump, collider and neutrino experiments on a kinetically mixed dark photon~\cite{Bauer:2018onh,Agrawal:2021dbo,CERN-FASER-CONF-2023-001}.

Collider searches for inelastic DM excited states (rather than the mediator) are possible~\cite{Izaguirre:2015zva, Berlin:2018jbm, CMS:2023ojs}; however, these searches have focused on dark photons with $m_{A'} > m_{\chi_1}+m_{\chi_2}$, as motivated by the larger production cross section and the benchmark of the visible (rather than dark sector) freeze-out thermal history. Such collider searches do not constrain the freeze-in scenario described in this work because we consider lighter mediators $m_{A'} < m_{\chi_1}+m_{\chi_2}$, the production cross sections are suppressed by the tiny freeze-in couplings, and the excited state is cosmologically long-lived, beyond the range of timescales that can be probed at colliders.

We also note that there may be direct detection signatures of the cosmologically populated excited state (rather than being produced in a collider) if it has a lifetime that is longer than the age of the Universe.  This is possible in parts of the parameter space due to the small mass splitting and small freeze-in couplings, and is trivially true if the mass splitting drops below the $\sim$MeV two electron threshold. This would lead to a situation where the relic DM excited state can potentially down-scatter in terrestrial experiments and deposit energy. Traditional direct detection experiments are sensitive to a range of energy depositions that are below the mass splittings we consider in this work, but neutrino experiments could have sensitivity in the appropriate energy range. Despite the large exposures provided by neutrino experiments, a preliminary estimate suggests that due to the very small couplings relevant for freeze-in, the expected number of observed down-scattering events is much less than unity. Therefore, we find no meaningful possibility of probing this parameter space via direct detection.

\subsection{Scattering processes}
\begin{figure}[t!]
\includegraphics[width=0.5\textwidth, trim={15 14 10 0}, clip]{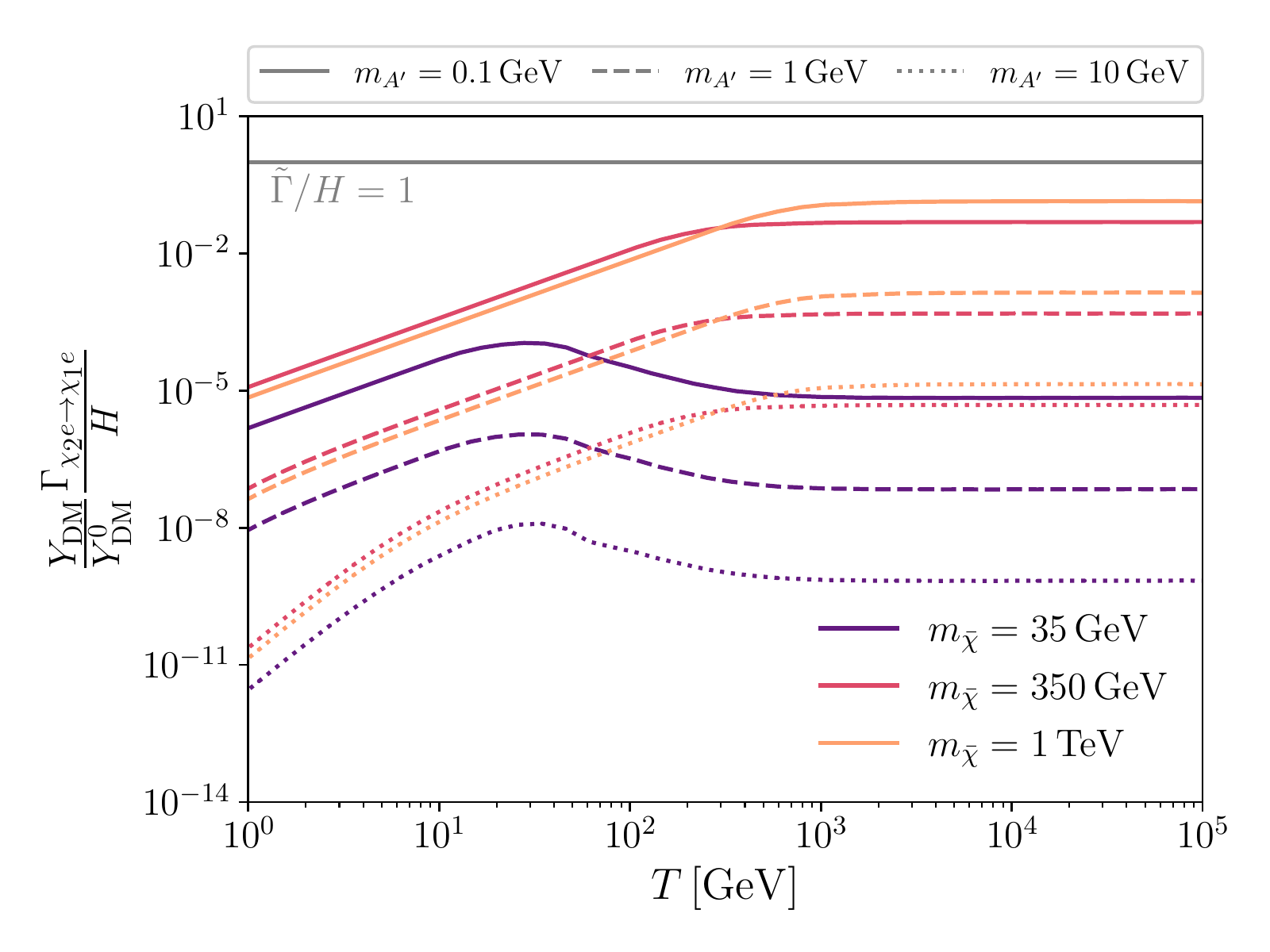}
\caption{\label{fig:coscattering} The normalised coscattering rate of DM with electrons, $\chi_1 e\leftrightarrow\chi_2 e$, compared to the Hubble rate as a function of temperature for different values of $m_{\bar{\chi}}$ and $m_{A^\prime}$. The couplings are chosen to satisfy the DM relic abundance. See text for details.}
\end{figure}

Although the ground and excited states are populated equally by freeze-in, scattering processes may change their relative number density after production. In this section, we consider processes that convert $\chi_2$ to $\chi_1$ (and vice versa).
If efficient, these reactions can establish chemical equilibrium in the dark sector with the ground and excited state densities related by $n_{\chi_2} \sim e^{-\delta/T_\chi} n_{\chi_1}$, where $T_\chi$ is the dark sector temperature. Accordingly, any interactions that keep the dark sector in chemical equilibrium can deplete the $\chi_2$ density for $T_\chi \lesssim \delta$~\cite{An:2020tcg,Baryakhtar:2020rwy}. We show below that these processes are negligible in our parameter space of interest.
\begin{itemize}
  \item Co-scattering off of SM fermions: $\chi_1 f \leftrightarrow \chi_2 f$. The dominant coscattering channel is with electrons. The rate of this process, $\Gamma = n_e\sigma v$, scales as $e^2 \epsilon^2 \cos^2\theta_W g_\chi^2 T^3/m_{A^\prime}^2$ in the high-temperature limit, and therefore could become efficient with $\Gamma >H$ at early times. However, at such large temperatures, the number densities of $\chi_{1,2}$ are negligible compared to the final freeze-in relic abundance. A better metric for comparison is therefore the rate normalised with the fractional comoving abundance of dark matter, $\tilde{\Gamma} = (Y_\mathrm{DM} (T)/Y^0_\mathrm{DM})\Gamma$, where $Y^0_\mathrm{DM}$ is the total comoving abundance of DM today. In Fig. ~\ref{fig:coscattering}, we plot the normalised coscattering rate of DM with electrons as a function of temperature for different values of $m_{\bar{\chi}},\, m_{A^\prime}$, and the couplings that produce the observed DM abundance. 
  The requirement that this rate be sub-Hubble at freeze-in, $T\sim m_{\bar \chi}$, results in a lower bound on the dark photon mass for a given DM mass. Using our estimate of $\Gamma$ gives a rough bound of $m_{A^\prime}\gtrsim e\epsilon \cos \theta_W g_\chi \sqrt{M_\mathrm{pl} m_{\bar\chi}}$ $\sim~0.29 \, {\rm GeV} \sqrt{m_{\bar\chi}/{\rm TeV}} $.
  To determine a more careful bound, as well as to show the rate in Fig. \ref{fig:coscattering}, we calculate the $\chi_1 f\to \chi_2 f$ cross-section numerically using \texttt{CalcHEP}. This gives $m_{A^\prime} > 28\, {\rm MeV}$ for $m_{\bar \chi} = 350$ GeV and $m_{A^\prime} > 48\, {\rm MeV}$ for $m_{\bar \chi} = 1$ TeV, somewhat lower than the analytic estimate.

  \item Up/down-scattering: $\chi_1\chi_1\leftrightarrow\chi_2\chi_2$. The cross-sections for this process have been studied in \cite{Schutz:2014nka} in the non-relativistic limit. In the parameter space we consider, $m_{\chi_1} \alpha_\chi \ll m_{A'}$ and therefore we can use the analytic results in the Born approximation derived in \cite{Schutz:2014nka}:
  \begin{equation}
  \label{eq:scatter_cs}
    \sigma_{\chi_1\chi_1 \to \chi_2\chi_2} = \frac{4\pi\alpha_\chi^2 m_{\chi_1}^2 \sqrt{1-\frac{2\delta}{m_{\chi_1} v^2}}}{m_{A^\prime}^4\left[\left(1-\frac{2\delta \, m_{\chi_1}}{m_{A^\prime}^2}\right)^2 +\frac{4 m_{\chi_1}^2v^2}{m_{A^\prime}^2} \right]}
    \end{equation}
    \begin{equation}
    \label{eq:scatter2_cs}
    \sigma_{\chi_2\chi_2 \to \chi_1\chi_1}
     = \frac{\sigma_{\chi_1\chi_1 \to \chi_2\chi_2}}{\left(1-\frac{2\delta}{m_{\chi_1} v^2}\right)}\,,
  \end{equation}
  where $v$ is the velocity of either of the DM particles in the center of mass frame. For DM produced via freeze-in, the momentum is inherited from the SM thermal bath with a typical momenta of $T$ and velocity~$v\sim T/m_{\chi}$. 
  Substituting for $v$ allows us to then calculate the corresponding rates, $n_{\chi_{1,2}} \sigma_{\chi_1\chi_1 \leftrightarrow\chi_2\chi_2}v$.
  The rate can also be analytically estimated as $\Gamma \sim 10^{-8}\, \mathrm{GeV} \times \alpha_\chi^2 T^4/(m_{A^\prime}^4 + 4 m_{A^\prime}^2 T^2)$, where the DM number density is obtained by scaling back its present day abundance, $\rho_\mathrm{DM}(T) \sim \rho_\mathrm{DM}^0 (T/T_0)^3$. For DM in our mass range, this corresponds to an upper bound on $\alpha_\chi \lesssim 10^{-7}$ at the temperatures relevant for freeze-in. As mentioned above, we are interested in the regime $g_\chi \lesssim e\epsilon \sim 10^{-6}$, and therefore in our parameter space of interest, these scattering processes are always sub-Hubble. We validate the above estimates with numerical calculations of the full cross-section using \texttt{CalcHEP}, since for $T \gtrsim m_{\bar{\chi}}$ the approximations in \cite{Schutz:2014nka} are no longer valid. The full results in the non-relativistic limit are closely approximated by the expressions given in Eqs. \ref{eq:scatter_cs} and \ref{eq:scatter2_cs}. Note that in our model, up- and down-scattering can also take place through a $Z$ in the $t$-channel, and we have checked numerically that the contribution from this diagram is negligible.
\end{itemize}

In summary, the dark sector never enters into chemical or kinetic equilibrium with itself or with the SM. Therefore the comoving number densities of the ground and excited states are the same as their values at the end of freeze-in production. The fact that $50\%$ of DM is present in the excited state until later times is necessary for having cosmological signatures. 

\section{Out-of-equilibrium decays}
\label{sec:decay}

Due to the small freeze-in couplings required to yield the observed DM relic abundance, there is no substantial DM annihilation rate to the SM via processes such as $\chi_2\chi_1 \to f \bar{f}$. At the time of recombination, this rate is well below CMB bounds on DM annihilation~\cite{Slatyer:2015jla, Planck:2018vyg}. However, the small freeze-in couplings can lead to a long lifetime for the excited DM state, with the result that 50\% of the DM is decaying in the late universe. The signatures of the late decay depend on the channel as well as the kinematics.

If $\delta < 2 m_e$, the excited state can only decay into $\chi_1 \nu \bar{\nu}$ and $\chi_1+3\gamma$. The rates for these processes are suppressed by a factor of $G_F^2$ and an electron loop respectively, as well as phase space factors, making the excited state extremely long-lived on cosmological timescales (see Appendix B1 of \cite{Fitzpatrick:2021cij}). On the other hand, for $\delta > 2 m_e$, the excited state can decay into charged leptons, $\chi_2\to\chi_1 \ell^+\ell^-$ with the dominant decay channels being electrons ($\delta > 2 m_e$) and muons ($\delta > 2 m_\mu$) for MeV-GeV scale $A'$. To a very good approximation, above the decay threshold the partial decay width is independent of the lepton mass and is given by,
\begin{align}
   \Gamma_{\chi_2 \to \chi_1 \ell^+\ell^-} \approx \frac{g_\chi^2 (e\epsilon \cos\theta_W)^2}{60\pi^3}\,\frac{\delta^5}{m_{A^\prime}^4}\,,
   \label{eq:chi2decayrate}
\end{align}
corresponding to a decay timescale of
\begin{align}
  \tau \sim 1.3 \times 10^7 \mathrm{s}\,\left(\frac{\epsilon g_\chi}{10^{-11}}\right)^{-2}\left(\frac{\delta}{100\,\mathrm{MeV}}\right)^{-5}\left(\frac{1\,\mathrm{GeV}}{m_{A^\prime}}\right)^{-4}\,.
\end{align}
This decay injects energetic leptons into the SM plasma and provides a velocity kick to the DM at times that are relevant to cosmological observables. Depending on the phase space of the decay products and the decay time, different cosmological probes can be used to hunt for signatures of this model. 

In order to calculate the phase space, we start by assuming that the relic population of $\chi_2$ is effectively at rest during the epochs relevant for cosmological observables. The consistency of this assumption is validated below. The phase space for the three-body decay is given by 
\begin{align}
\label{eq:dGammafull}
  \int\!\mathrm{d}\Gamma \!=& \frac{1}{m_{\chi_2}}\!\int\!\frac{\dbar^3p_{\chi_1}}{2E_{\chi_1}}\!\frac{\dbar^3p_{\ell^+}}{2E_{2}}\!\frac{\dbar^3p_{\ell^-}}{2E_{3}} |M|^2_{\chi_2 \to\chi_1 \ell^+ \ell^-}\nonumber \\
  & \times (2\pi)^4 \delta^4(p_{\chi_2}\! -\! p_{\chi_1}\!\! -\! \!p_{\ell^+}\! \!-\!\! p_{\ell^-})\,, 
\end{align}
where $\dbar^3 p_i = \mathrm{d}^3 \vec p_i/(2\pi)^3$ and the matrix element for the decay is
\begin{align}
  &|M|^2_{\chi_2 \to\chi_1 \ell^+ \ell^-}\, = \frac{(e\epsilon g_\chi \cos\theta_W)^2}{((p_{\chi_2} - p_{\chi_1})^2 - m_{A^\prime}^2)^2}\bigg[m_\ell^2 \,(p_{\chi_2}\!\cdot p_{\chi_1}) \nonumber \\
  &- m_{\chi_1}m_{\chi_2}(p_{\ell^+}\!\cdot p_{\ell^-}) + (p_{\chi_1}\!\cdot p_{\ell^-})(p_{\chi_2}\!\cdot p_{\ell^+}) \nonumber \\
  &+(p_{\chi_2}\!\cdot p_{\ell^-})
  (p_{\chi_1}\!\cdot p_{\ell^+}) -2m_\ell^2 m_{\chi_1} m_{\chi_2}\bigg].
\end{align}

In the rest frame of $\chi_2$, energy-momentum conversation dictates that the momentum of $\chi_1$ is equal and opposite to the net momentum of the $\ell^+\ell^-$ pair,
\begin{align}
|\vec{p}_{\chi_1}| 
=\frac{m_{\chi_2}}{2}\sqrt{1 \!-\! \frac{2(m_{\chi_1}^2\! +\! s_{23})}{m_{\chi_2}^2} \!+\! \frac{(m_{\chi_1}^2\!-\!s_{23})^2}{m_{\chi_2}^4}}\,,
\end{align}
where $\sqrt{s_{23}}$ is the effective mass of the lepton pair, $\sqrt{s_{23}} = \sqrt{E_{\ell^+\ell^-}^2\!-\! \vec p_{\chi_1}^{\, 2}}$ for net lepton pair momentum $\vec{p}_{\ell^+\ell^-} = - \vec p_{\chi_1}$. Then, using the standard treatment for a three-body phase space \cite{3body}, we can write the partial decay width as 
\begin{align}
\label{eq:dGammads}
\int\!\mathrm{d}\Gamma \!= \frac{1}{64\pi^2m_{\chi_2}}\!\int &\frac{d s_{23}}{2\pi} \!\sqrt{1\!-\!\frac{2(m_{\chi_1}^2\! +\! s_{23})}{m_{\chi_2}^2}\! +\! \frac{(m_{\chi_1}^2 \!-\! s_{23})^2}{m_{\chi_2}^4}}\nonumber \\
&\times\sqrt{1\!-\!\frac{4m_\ell^2}{s_{23}}}|\tilde{M} (s_{23})|^2\,.
\end{align}
Here $|\tilde M|^2$ is the matrix element squared integrated over all angles in the rest frame of $s_{23}$. Eq.~\ref{eq:dGammads} can then be used to obtain a differential decay rate in either $|\vec p_{\chi_1}|$ or in $E_{\ell^+\ell^-}$ using the relationships with $s_{23}$ above.

\begin{figure}[t]  
\includegraphics[width=0.49\textwidth]{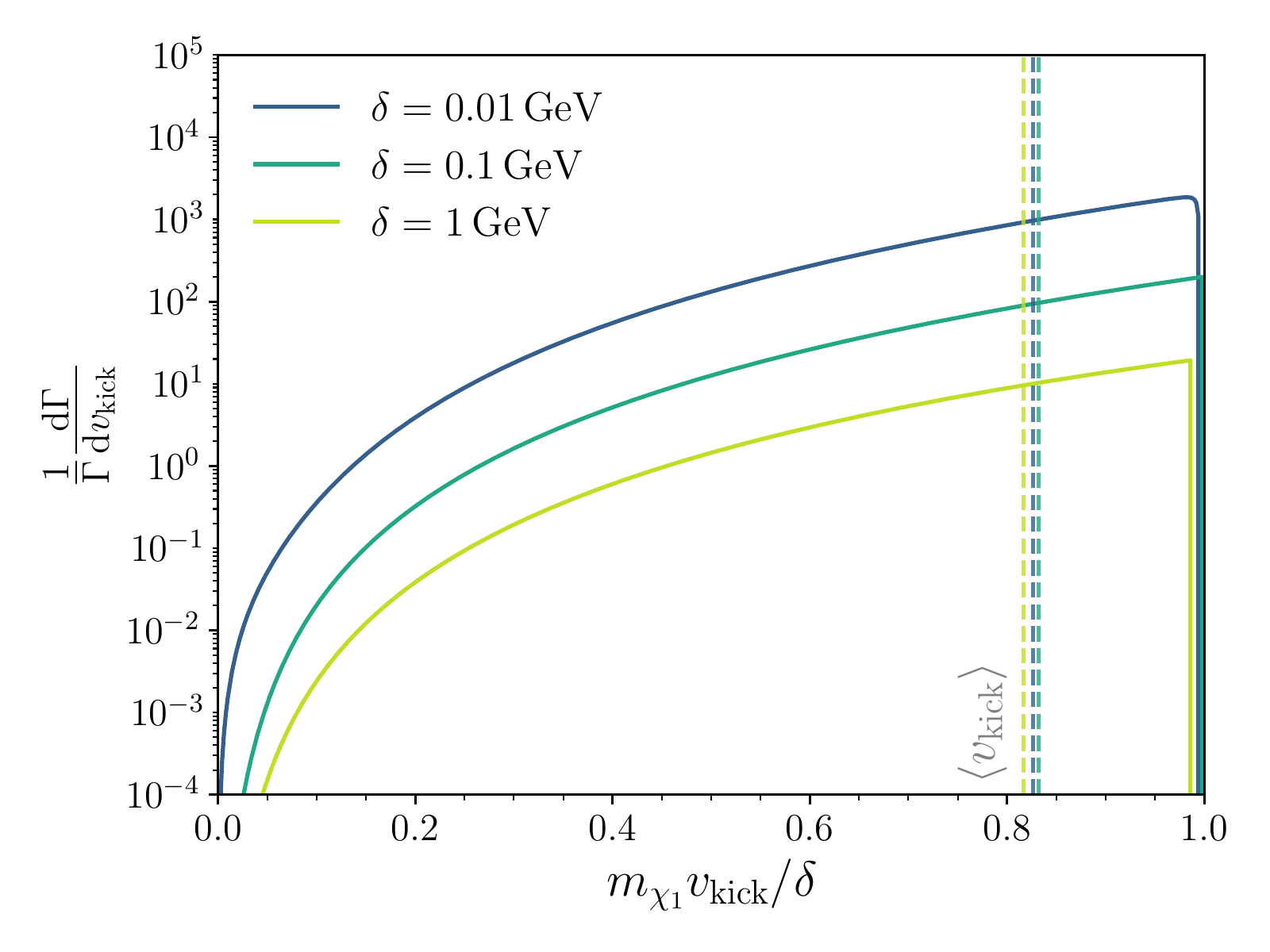}
  \caption{The differential decay width as a function of a normalised kick velocity, $m_{\chi_1}v_{\rm kick}/\delta$, for different values of $\delta$. For concreteness, lines are shown with $m_{\chi_1} = 35$ GeV; results with other $m_{\chi_1}$ are very similar. The dashed lines correspond to the average kick velocity, $\langle v_\mathrm{kick} \rangle$. 
  \label{fig:three_body}
  }
\end{figure}

We can parameterise the effect of this decay on the dark and visible sectors in terms of the average kick velocity to the $\chi_1$ produced in the decay, $\langle v_\mathrm{kick}\rangle$, and the average energy carried by the lepton pair, $\langle E_{\ell^+\ell^-} \rangle$. Since $s_{23}\!\in\! [4m_\ell^2, \,\delta^2]$, 
\begin{align}
\label{eq:px1lim}
  &0\leq |\vec{p}_{\chi_1}|\leq \frac{1}{2m_{\chi_2}}\sqrt{(\delta^2\!-\!4m_\ell^2)((m_{\chi_1}\!+\!m_{\chi_2})^2\! - \!4m_\ell^2)} \\
  \label{eq:Eelim}
  &\frac{\delta(m_{\chi_1}+m_{\chi_2}) + 4m_\ell^2}{2m_{\chi_2}} \leq E_{\ell^+\ell^-} \leq \delta\,.
\end{align}
In Fig.~\ref{fig:three_body}, we plot the normalised differential decay width as a function of the velocity kick given to $\chi_1$ for $m_{\chi_1} = 35\,\mathrm{GeV}$, $m_{A^\prime} = m_{\chi_1}/10$ and different values of $\delta$. The dashed lines represent the average value of the kick velocity. Note that although we plot the curves for specific values of $m_{\chi_1}$ and $m_{A^\prime}$, the normalisation of the axes ensures that, for a given $\delta$, any value of $m_{\chi_1}, m_{A^\prime}$ will essentially map on to the same lines. Additionally, Eq.~\ref{eq:Eelim} implies that the energy transferred to the leptons is sharply peaked around the mass splitting. 

It follows that the average kick velocity and the average energy transferred are given by, 
\begin{align}
\label{eq:vkick}
\langle v_\mathrm{kick}\rangle \approx \frac{\delta}{m_{\chi_1}}\,, \quad
\langle E_{\ell^+\ell^-} \rangle \approx \delta\,.
\end{align}
These values are derived in the rest frame of $\chi_2$, and will be essentially the same in the comoving frame if the typical speed of $\chi_2$ is much smaller than $v_{\rm kick}$ at the time of decay. Freeze-in ends when the SM temperature $T \sim m_{\bar \chi}$ and the $\chi_2$ has a typical momentum of $p_{\chi_2} \sim T$ at this time. Since the DM is not in kinetic equilibrium, the momentum redshifts along with the SM thermal bath temperature and at late times $v_{\chi_2} \sim T/m_{\chi_2}$. This residual speed at the time of decay will be smaller than $v_{\rm kick}$ as long as $T \ll \delta$, which is a valid approximation for the cosmological observables that we consider in this work. We can now apply the distributions derived above to a range of interesting consequences for cosmological observables in the following Section.

\section{Cosmological Signatures}
\label{sec:cosmo}
The out-of-equilibrium decays of the metastable excited state can have observable implications in ongoing and future cosmological probes. In this section, we focus on the two complementary classes of signatures coming from (1) energy injection into the baryonic sector and (2) suppressing structure formation on small scales due to streaming effects. 

\subsection{Energy injection from charged lepton pairs}

\begin{figure*}
  \centering
\includegraphics[width=\textwidth]{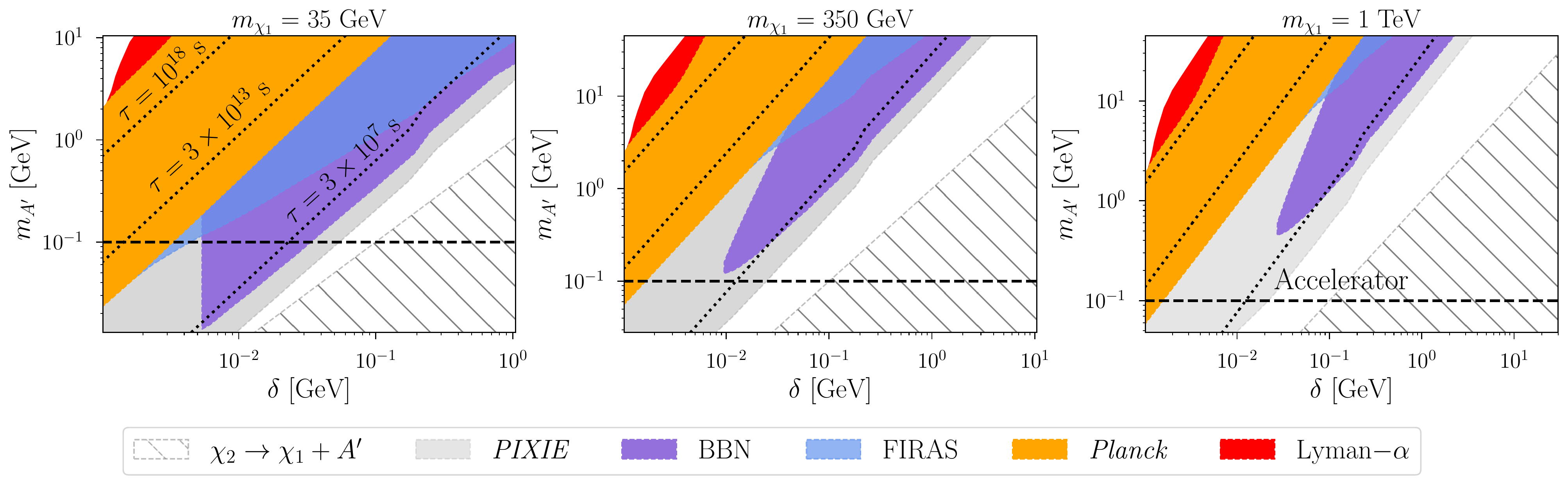}
  \caption{Constraints and forecast sensitivity for the dark photon mass and DM mass splitting at different DM masses corresponding to $m_{\chi_1}= $~35~GeV, 350~GeV, and 1~TeV (left, middle, and right panels). For all constraints, it is assumed that the coupling combination $g_\chi \epsilon$ is fixed by saturating the DM relic abundance with freeze-in. In the hatched grey region, the excited state is not cosmologically relevant because it decays extremely rapidly (right after freeze-in) via $\chi_2 \rightarrow A' \chi_1$. The mass splitting is shown for values above the $\sim$MeV electron production threshold for three-body decay. Meanwhile, the depicted dark photon masses range from the minimum allowed by beam dump and collider experiments (shown as a black $\times$ in the left panel of Fig.~\ref{fig:acc_bounds}) or alternatively the minimum $m_{A'}$ from becoming cosmologically relevant as shown in Fig.~\ref{fig:coscattering}. The maximum value of $m_{A'}$ depicted goes up to $m_{A'} = m_{\chi_1}/3$, since we focus on the parameter region with $m_{A'} \ll m_{\chi_1}$, but for heavier DM masses we also ensure that $m_{A'} \ll m_{Z}$ which was an assumption made in the freeze-in analysis. The dashed horizontal line represents the minimum $m_{A'}$ that can be accessed in the near future with beam dump and collider experiments in the window of visible freeze-in couplings, shown as a black $\times$ in the right panel of Fig.~\ref{fig:acc_bounds}. Meanwhile, the diagonal dotted lines correspond to contours of constant $\chi_2$ lifetime for a few representative values (corresponding roughly to (i) the age of the Universe, (ii) shortly after recombination, and (iii) one year after the Big Bang, from left to right). 
  \label{fig:inj}}
\end{figure*}
As shown in the previous section, the $e^+ e^-$ pairs (or $\mu^+ \mu^-$ pairs, if the mass splitting is above threshold) from the three-body decay get an $\mathcal{O}(1)$ fraction of the energy available from the mass splitting. These charged particles can subsequently deposit their energy into the photon-baryon plasma or intergalactic medium (IGM), causing ionization and heating. The efficiency of these effects depends on the energy of the electrons (or other SM particles) and ambient properties of the plasma or IGM, and is captured via a parameter $f_\text{eff}(z)$, defined as the ratio between the rate of injected and deposited energy density,
\begin{align}
\frac{d E_\text{dep.}}{dV \, dt}&= f_\text{eff}(z) \frac{d E_\text{inj.}}{dV \, dt} \\
  &= f_\text{eff}(z) \frac{ \rho_{\rm DM}(z) \, \delta}{2 m_{\bar \chi}} \Gamma e^{- \Gamma t}. \label{eq:feff}
\end{align}
In the second line, we have used the fact that half of the DM density is initially in the excited state and essentially all of the energy from the mass splitting goes into the leptons. Here $\Gamma = \Gamma_{\chi_2 \to \chi_1 \ell^+ \ell^-}$. The efficiency $f_\text{eff}(z)$ depends on the redshift and energy spectrum of injected particles. Depending on the redshift when the decay occurs, this energy deposition can alter BBN, the spectrum of cosmic microwave background (CMB) photons, CMB anisotropies, and Lyman-$\alpha$ forest measurements. There is also a possibility for 21~cm signatures in the future~\cite{Furlanetto:2006wp, Finkbeiner:2008gw, Evoli:2014pva, Liu:2018uzy}, pending an improved understanding of the relevant baryonic astrophysics and observational systematics. Assuming coupling constants fixed by freeze-in to produce the DM relic abundance, we can then place bounds in the parameter space of $m_{\bar \chi}, m_{A'}$, and $\delta$. All of the bounds and forecast sensitivities are summarized in Fig.~\ref{fig:inj}. 

For decay lifetimes $\tau \lesssim 10^{12}$~s, BBN currently provides the tightest constraints on decaying DM~\cite{Kawasaki:2004qu,Poulin:2015opa,Poulin:2016anj,Forestell:2018txr}. At times prior to recombination, energy injected in the form of $e^+e^-$ pairs is approximately instantly deposited into the plasma and $f_\text{eff} \approx 1$. The energy injection from DM three-body decays can therefore enhance the dissociation of light elements produced during BBN. The yields of these elements are very sensitive to changes in the ambient environment during and after BBN, and thus the agreement of the measured yields with the $\Lambda$CDM prediction can be used to constrain DM decay at early times. 
Note that the dissociation only proceeds if the spectrum of injected electron energies has support above the $\sim3$~MeV energy threshold for breaking apart light nuclei. For our model, the {\it combined} electron-positron energy is effectively monochromatic, but the energy of each of the individual particles can vary. This $\sim3$~MeV threshold can have a large impact on the bounds when $\delta$ is near the MeV-scale. For instance, Ref.~\cite{Poulin:2016anj} bounds the total fraction of DM energy injected into the electromagnetic sector assuming electrons are injected with energy above 10 MeV, while Ref.~\cite{Forestell:2018txr} finds that there is a large impact on the limits when the energy of the electrons injected is at the $\sim$10 MeV scale. Without doing a re-analysis of BBN bounds with our specific electron spectrum, we conservatively apply the BBN constraints of Ref.~\cite{Poulin:2016anj} when $\delta>6$~MeV and note that it is possible the bound extends slightly below this. The application of this cut is apparent in the left panel of Fig~\ref{fig:inj}, where the shape of the BBN bound differs from the other panels. In all, we find that due to the dependence of the decay rate on $\delta$ and $m_{A'}$, BBN is currently able to exclude a large swatch of parameter space for pseudo-Dirac freeze-in. 

When the decay occurs with a range of decay lifetimes $10^7 \text{~s} \lesssim \tau \lesssim 10^{13}$~s, the energy injection heats the plasma which alters the spectrum of CMB photons. The spectrum does not have time to re-thermalize if the energy injection happens below redshift $z\sim 10^6$, corresponding to $\tau \sim 10^7$~s or around 1~year after the Big Bang. This causes a spectral distortion where the CMB spectrum deviates from a perfect blackbody~\cite{Chluba:2011hw}. During the spectral distortion window, charged particles deposit their energy into the plasma almost instantaneously in a way that is insensitive to the energy spectrum of injected electrons, so again $f_\text{eff} \approx 1$ with no strong dependence on $\delta$. 

There are strong bounds on deviations of the CMB spectrum from a perfect blackbody, which can then be translated to bounds on a fraction of the DM decaying to energetic electrons~\cite{Lucca:2019rxf}. These bounds as applied to our parameter space are shown in Fig.~\ref{fig:inj} in the region labelled ``FIRAS,'' in reference to the Far Infrared Absolute Spectrometer aboard the Cosmic Background Explorer (COBE)~\cite{Fixsen:1996nj}. While current spectral distortion limits are not very competitive with other probes in the relevant decay lifetime window (almost everything excluded by FIRAS is already excluded by BBN), there would be a large improvement in sensitivity to our parameter space~\cite{Lucca:2019rxf, Chluba:2013pya} with the proposed Primordial Inflation Explorer (PIXIE) mission~\cite{Kogut:2011xw}, as shown in the light grey region in Fig.~\ref{fig:inj} labelled ``{PIXIE}.''

At times during and after recombination, $\tau \gtrsim 10^{13}$~s, the injection of energetic electrons can partly ionize and heat the IGM. In $\Lambda$CDM, the IGM is almost completely neutral after recombination and the photons free-stream to good approximation; ionization of the IGM due to DM decay will cause CMB photons to re-scatter, smearing out the anisotropies. Strong bounds on the ionization fraction of the IGM from {Planck} anisotropy measurements constrain the deposition of ionization energy, which can be translated to a constraint on decaying DM. Similarly, there is a possibility that energy injection will heat the IGM during epochs around $z\sim 5$, where the Lyman-$\alpha$ forest of atomic absorption lines constrains the IGM temperature. Given a measurement of this temperature, one can exclude processes which would further heat the IGM, allowing a constraint to be set on the DM decay lifetime. For both ionization and heating, $f_\text{eff}$ can vary substantially as a function of the injected energy spectrum, species, and redshift of injection, requiring a careful analysis involving the use of specialized codes which self-consistently take into account the effects of DM decay on the thermal history of the IGM, including backreactions~\cite{Stocker:2018avm,Liu:2019bbm}. We leave such an analysis as applied to our model for future work, and note that because our charged particle spectra are not mono-energetic, we are only able to estimate the corresponding limits.

In order to estimate the limits on our parameter space, for decay lifetimes near the time of recombination we apply the bounds of Ref.~\cite{Lucca:2019rxf}. This work constrained the fraction of DM decaying as a function of decay lifetime assuming $f_\text{eff}= 1$, which is a good approximation during and before recombination. When translated to our parameter space we find that the resulting limit is not very sensitive to this assumption about $f_\text{eff}$. At late times well after recombination, however, $f_\text{eff}$ can vary considerably so we apply
the bounds from Ref.~\cite{Slatyer:2016qyl} for CMB anisotropies and Ref.~\cite{Liu:2020wqz} for the Lyman-$\alpha$ forest which solved for the deposition efficiency in a self-consistent way. Instead of constraining the fraction of DM decaying as a function of decay lifetime, as was done in Ref.~\cite{Lucca:2019rxf}, these works instead constrain the DM decay lifetime as a function of DM mass, and the effects of the energy spectrum on $f_\text{eff}$ cause this limit to vary considerably for different DM masses. Both of those works considered DM decay where the electron energy injection spectrum is centered at half the DM mass, so we translate their bounds by matching their DM mass with our mass splitting, noting that this translation is only an estimate because our electron energy spectrum is broad while theirs is effectively a mono-energetic line. Moreover, the energy density available from the DM mass splitting is a small fraction of the total energy density of the DM. We therefore estimate the constraint on our parameter space by re-scaling the decay rate to account for the lower energy density available in the mass splitting so that the {\it total} energy injected into the IGM is the same. Concretely, we re-scale the lifetime bounds of Refs.~\cite{Slatyer:2016qyl} and \cite{Liu:2019bbm} by $\delta/2 m_{\bar{\chi}}$, as in Eq.~\eqref{eq:feff}. This re-scaling is supported by works that consider a sub-component of DM decaying rather than looking at all of the DM having a particular decay lifetime~\cite{Slatyer:2016qyl, Poulin:2016anj, Lucca:2019rxf}. The resulting estimated limits on our parameter space, combining early and late times, are shown as shaded regions in Fig.~\ref{fig:inj}.  

For lifetimes comparable to the age of the Universe, there can also be constraints on decaying DM from indirect searches for the decay products, e.g. with X-ray and gamma ray searches~\cite{Essig:2013goa, Cohen:2016uyg,Massari:2015xea,Cirelli:2023tnx} which generally are weaker than cosmological constraints from the IGM in the parameter space of interest. One exception to this is the search for cosmic ray electrons and positrons directly from the decay; these particles cannot be detected from inside the solar system due to shielding from the solar magnetic field, but they can be observed by Voyager I~\cite{Cummings:2016pdr}, which has crossed the heliopause. The resulting limits on decaying DM of Ref.~\cite{Boudaud:2016mos} are stronger than cosmological ones for mass splittings greater than $\delta \gtrsim20$~keV, however when the corresponding lifetimes are mapped onto our parameter space, the values of $m_{A'}$ are too large for the analysis in this work to be valid. Therefore, these limits do not appear in Fig.~\ref{fig:inj} as the relevant parameter space probed by indirect detection searches is already excluded by cosmological ones.

\subsection{Free-streaming of the ground state}

\begin{figure*}
  \centering
\includegraphics[width=\textwidth]{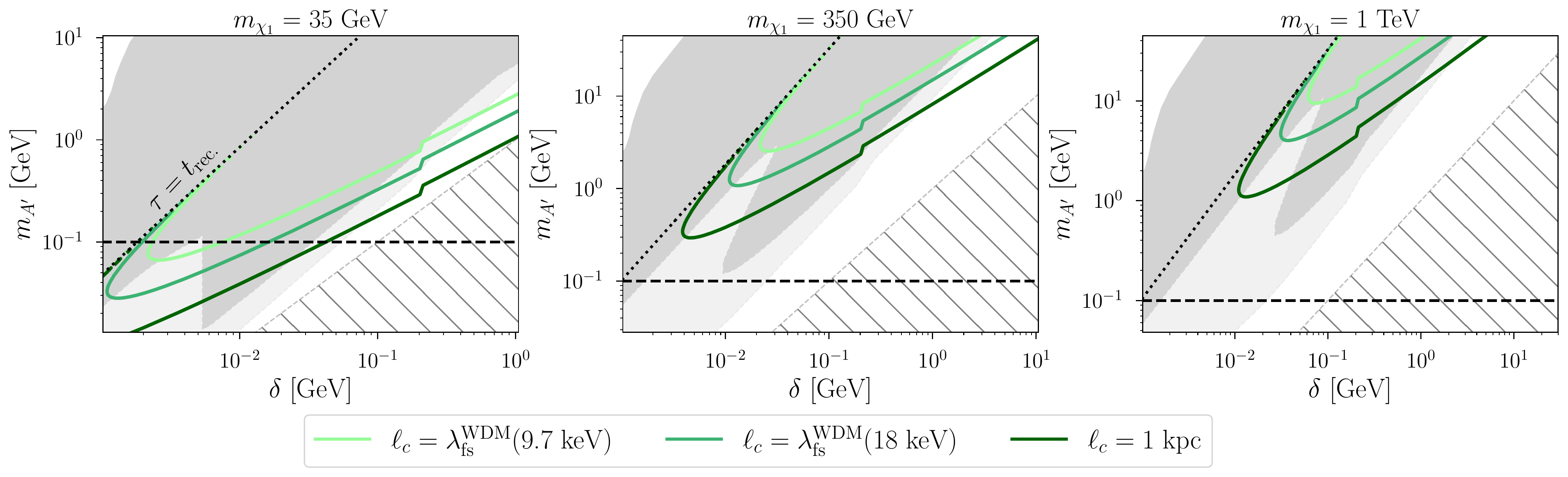}
  \caption{The same parameter space as Fig.~\ref{fig:inj} but showing the complementarity with probes of structure formation. Existing cosmological constraints from Fig.~\ref{fig:inj} are shown here as dark grey regions, while the forecast for PIXIE is shown in light grey. The green lines correspond to parameters that give a decay timescale and velocity kick that will yield a free streaming length $\ell_c(t_\text{rec})$, as estimated in Eq.~\eqref{eq:kick}, equivalent to that of WDM at a given mass.}
  \label{fig:kickparam}
\end{figure*}

As shown in the previous section, as the relic 50\% of DM in the excited state decays, the resulting ground state daughter particle receives a velocity kick with a recoil spectrum peaked near $v \sim \delta/m_\chi$. As a result, the 50\% of the DM that was born in the ground state behaves as cold DM, while the other 50\% that gets produced from the decay can undergo some free streaming, similar to warm DM (WDM) or neutrinos. This effect has been previously studied in detail in situations where all of the DM decays via a two-body process to the ground state and a light degree of freedom~\cite{Wang:2012eka,Wang:2013rha}. In this section, we instead estimate the free-streaming length for the ground state population produced via three-body decays, with a more detailed study of structure formation in this model left to future work. Based on our estimates, we find untested parameter space to be within the reach of current probes of structure formation on small scales, and potentially of interest for future probes.

To estimate the scale of suppression of the transfer function relevant for the matter power spectrum, we employ the commonly used metric of the free-streaming length. We estimate this by assuming that the DM gets a physical velocity kick $v_{\rm kick}$ at a time set by the lifetime $\tau$ and that it free-streams with a Hubble-damped velocity. Ignoring any motion prior to the kick, the co-moving distance travelled by the DM by some later time $t$ is 
\begin{equation}  
\ell_c(t) = \int_\tau^t \frac{v_d a(\tau) dt'}{a(t')^2} = \int_{z(t)}^{z(\tau)} \frac{v_d (1+z) dz}{H(z) (1+ z(\tau))} \label{eq:kick}
  \end{equation}
where $H(z) = H_0 \sqrt{\Omega_\Lambda + \Omega_m (1+z)^3 + \Omega_r (1+z)^4}$. For the energy densities and present-day Hubble parameter, we take the best-fit values from {Planck}~2018~\cite{Planck:2018vyg}. 
Note that the approximation of free-streaming is not a good one for times long after matter-radiation equality (i.e. deep in the matter-dominated era) because of the onset of structure formation leading to gradients in the gravitational potential. In order to be conservative in our estimate of the co-moving free streaming length today, we consider the free-streaming length at the time of recombination, $\ell_c(t_\text{rec})$. This means that our estimate only applies in the parameter space to the right of the contour of constant $\tau = t_\text{rec}$, as shown in Fig.~\ref{fig:kickparam}.

In order to estimate the sensitivity of probes of structure formation in our parameter space, we equate $\ell_c(t_\text{rec})$ with the free-streaming length of the well-studied WDM model, $\lambda_\text{fs}^\text{WDM} = 0.07 \times (m_\text{WDM} / \text{1 keV})^{-1.11}$~Mpc for a Hubble parameter of $H_0=70$~km/s/Mpc~\cite{bode2001halo,viel2005constraining,Schneider:2011yu}. This free-streaming length maps onto the transfer function for the matter power spectrum, as determined via Boltzmann solvers, and leaves an imprint on the subsequent formation of structure on small scales. There are many probes of structure on small scales in linear and nonlinear regimes, including the Lyman-$\alpha$ forest~\cite{Irsic:2017ixq,Murgia:2018now, Rogers:2020ltq}, the detection of low-mass subhalos with galaxy surveys~\cite{Nadler:2019zrb,DES:2020fxi}, gravitational lensing and perturbations to stellar streams induced by subhalos~\cite{hsueh2020sharp, Gilman:2019nap,Banik:2019cza,Banik:2019smi}, and eventually 21~cm cosmology~\cite{Munoz:2019hjh}. Combining these probes can also help to break degeneracies and strengthen bounds~\cite{Enzi:2020ieg,Nadler:2021dft}; in particular, Ref.~\cite{Nadler:2021dft} was able to set a relatively aggressive bound of $m_\text{WDM}>9.7$~keV at 95\% confidence.

In Fig.~\ref{fig:kickparam} we show the parameter points that yield an estimated free-streaming length for pseudo-Dirac freeze-in which matches that of a 9.7~keV WDM model. In the near future, the Vera Rubin Observatory is set to target $m_\text{WDM}\sim 18$~keV by probing the subhalo mass function down to masses of $10^6 M_\odot$ through a combination of the effects of subhalos mentioned above~\cite{Drlica-Wagner:2019xan}; we therefore also show the freeze-in parameters that match the corresponding WDM free-streaming length in Fig.~\ref{fig:kickparam}. Finally, more futuristic probes of structure formation on small scales could probe free-streaming lengths well beyond these estimates~\cite{Bechtol:2022koa}; we show the parameter space that could be accessed with sensitivity to free-streaming lengths at the kpc scale. Overall, the structure formation signatures of this model are manifest in a complementary portion of the parameter space compared to other cosmological probes as well as terrestrial signatures.

We note that DM that decays to a warm lighter state plus relativistic species (like highly boosted electrons) can affect the expansion history of the Universe and potentially alleviate the Hubble and S8 tensions~\cite{Blackadder:2015uta,Vattis:2019efj,Haridasu:2020xaa,FrancoAbellan:2020xnr,FrancoAbellan:2021sxk}. However, decay lifetimes of interest for these tensions are generally comparable to the age of the Universe at recombination. We find that the parameter space that could observably alter the expansion history is excluded by strong CMB constraints on injected energetic electrons, as shown in Fig.~\ref{fig:inj}.

We additionally note that the presence of the excited state can have an \emph{in situ} impact on the internal dynamics of DM halos after they have already collapsed. Due to the small couplings relevant for freeze-in (and especially the small values of $g_\chi$ considered in this work), DM self-scattering is unlikely to be efficient enough for the inelastic nature of the collisions to alter halo properties as in Refs.~\cite{Vogelsberger:2018bok,Chua:2020svq,ONeil:2022szc}. Meanwhile, the three-body decay of the excited state could impact the halo through the velocity kick imparted to the ground state. The resulting bounds on the decay lifetime are generally comparable to the age of the Universe, ruling out $\tau \lesssim 30$~Gyr under the assumption that the lifetime is sufficiently long to only occur only after halos have formed~\cite{Peter:2010jy,Wang:2014ina,DES:2022doi}. For this range of lifetimes, the CMB constraints on the injection of energetic electrons already exclude the whole parameter space, as shown in Fig.~\ref{fig:inj}.

\section{Discussion and Conclusions}
\label{sec:conclusions}

In this work, we have explored the freeze-in regime of pseudo-Dirac DM coupled to the SM via a kinetically mixed dark photon. For DM in the mass range $\sim 30\, {\rm MeV} - {\rm TeV}$ with mass splittings above an MeV, we showed that there are complementary probes of this model from accelerator searches for dark photons and cosmological observables. For dark photon couplings accessible with near-future accelerator searches, the primary processes determining the thermal history are the freeze-in production channels $ f \bar{f} \rightarrow \chi_1 \chi_2$ and $Z\rightarrow \chi_1 \chi_2$, with a negligible amount of scattering through other channels. 

If the mass splitting is above $\sim1$~MeV then the excited state can decay to the ground state plus a pair of electrons (or muons at higher mass splittings). The decays can inject energy into the ambient environment, affecting BBN, CMB spectral distortions and anisotropies, and the Lyman-$\alpha$ forest. We find that the freeze-in mechanism
can be realized in a broad region of parameter space that
is not excluded by current constraints. We additionally find that much of the currently allowed parameter space would be explored with a future mission like PIXIE. Finally, we explored the possibility of a cosmological signature arising due to the kick imparted to the ground state DM particle during the three-body decay, which could impact structure formation on small cosmological scales that will be explored in the near future.

\section*{Acknowledgements}
It is a pleasure to thank Asher Berlin, Xiaoyong Chu, Patrick Foldenauer, Felix Kahlhoefer, Wenzer Qin, and Tracy Slatyer for useful conversations and correspondence pertaining to this work. SH acknowledges fellowship funding from the Trottier Space Institute at McGill. SH and KS acknowledge support from a Natural Sciences and Engineering Research Council of Canada (NSERC) Subatomic Physics Discovery Grant. TL acknowledges support from the US Department of Energy under grant DE-SC0022104.
\bibliography{manuscript}
\end{document}